\documentclass{article} 

\pdfoutput=1

\usepackage{arxiv}

\usepackage{url}
\usepackage{todonotes}
\usepackage{hyperref}
\usepackage{listings}
\usepackage{xspace}
\usepackage{MnSymbol}
\usepackage{paralist}
\usepackage{amsmath}
\usepackage{floatrow}

\newcommand{\eg}{e.g.\xspace}
\newcommand{\ie}{i.e.\xspace}

\newcommand{\toolname}[1]{\textsl{#1}\xspace}

\renewcommand{\shorttitle}{On the Security Blind Spots of Software Composition Analysis}

\title{On the Security Blind Spots of Software Composition Analysis caused By Cloning and Shading }

\date{June 7, 2023}

\author{
	Jens Dietrich \\
	Victoria University of Wellington \\
	Wellington, New Zealand \\
	\texttt{jens.dietrich@wgtn.ac.nz} \\
	\And
	Shawn Rasheed \\
	UCOL | Te Pūkenga \\
    Palmerston North, New Zealand \\
    \texttt{unshorn@gmail.com} \\		
    \And
    Alexander Jordan \\
    Oracle Labs, Vienna \\
    \texttt{alexander.jordan@oracle.com} \\
    \And
    Tim White\\
	Victoria University of Wellington, Wellington \\ 
	New Zealand \\
    \texttt{tim.white@vuw.ac.nz} \\
} 

\hypersetup{
	pdftitle={On the Security Blind Spots of Software Composition Analysis caused By Cloning and Shading},
	pdfauthor={Jens Dietrich, Shawn Rasheed, Alexander Jordan},
	pdfkeywords={vulnerability detection, clone detection, shading, sofwtare composition analysis, Java, Maven},
}

\DeclareFloatFont{tablesize}{\scriptsize}

\begin{document}
\maketitle

\floatsetup[table]{font=tablesize}

%
%


\begin{abstract}

Modern software heavily relies on the use of components. Those components are usually published  in central repositories, and managed by build systems via dependencies. Due to issues around vulnerabilities, licenses and the propagation of bugs, the study of those dependencies is of utmost importance, and numerous software composition analysis tools have emerged for this purpose. A particular challenge are hidden dependencies that are the result of cloning or shading where code from a component is "inlined", and, in the case of shading,  moved to different namespaces. 

We present a novel approach to detect vulnerable clones in the Maven repository.  Our approach is lightweight in that it does not require the creation and maintenance of a custom index. Starting with 29 vulnerabilities with assigned CVEs and proof-of-vulnerability projects, we retrieve over 53k potential vulnerable clones from Maven Central. After running our analysis on this set, we detect 727 confirmed vulnerable clones (86 if versions are aggregated) and synthesize a testable proof-of-vulnerability project for each of those. We demonstrate that existing SCA tools often miss those exposures.  At the time of submission those results have led to changes to the entries for six CVEs in the GitHub Security Advisory Database (GHSA) via accepted pull requests, with more pending.

\end{abstract}

%


\maketitle

\section{Introduction and Background}
\label{sec:introduction}

Modern software systems often use components in order to achieve economy of scale. The process is recursive -- components also use other components, resulting in deep and complex component ecosystems~\cite{wittern2016look,kikas2017structure,decan2019empirical}. This has in turn created new challenges. The prime example is vulnerability propagation, infamous examples include the \textit{equifax}~\cite{CVE-2017-5638,luszcz2018apache} and \textit{log4shell}~\cite{CVE-2021-44228,hiesgen2022race} incidents, with vulnerable and outdated components now being acknowledged as being a major security risk~\cite{owaspTop10A06}. Other related issues include license compliance~\cite{riehle2019open}, typo-squatting~\cite{taylor2020defending}, and lifecycle issues of components as demonstrated by the  \textit{leftpad} incident~\cite{chowdhury2021untriviality}.

In response to those challenges, software composition analysis (SCA)  tools have emerged that scan the dependency networks, and cross-reference them with known vulnerabilities catalogued in databases such as the National Vulnerability Database (NVD)~\cite{nvd} and the GitHub Advisory Database~\cite{ghsa}. If a vulnerable dependency is found, developers are notified and can upgrade dependencies to a newer version.  Examples of such tools include GitHub's \textit{dependabot}~\cite{dependabot}, \textit{snyk}~\cite{snyk},
\textit{OWASP dependency check}~\cite{owasp-dependency-check},  tooling integrated into development environments such as \textit{IntelliJ's dependency analysis} backed by \textit{checkmarx},  and features or plugins of build tools like \textit{npm audit} (for JavaScript) and Sonatype's \textit{oss index} Maven plugin~\cite{sonartype-plugin}.

At a high-level an SCA tool combines two components, a scanning component to find dependencies, and a vulnerability database (\emph{vulnerability DB}) to decide whether a dependency has a known vulnerability or not.
To combine the two, some matching logic is required, which provides a bridge between the low-level packages used by build systems (e.g., Maven artifacts in the Java/Maven ecosystem) and a more coarse-grained software identifier at the product-level, like the CPE (Common Platform Enumeration) standard used by NVD.
This matching is not always straight-forward, it varies across tools, and can introduce inaccuracies. Versions are another source of inaccuracy for SCA tools, often due to the fact that it is hard to pinpoint when a vulnerability was introduced, in which case (conservative) assumptions have to be made.

With the exception of \toolname{Eclipse Steady}, which uses program analysis to determine reachability of vulnerable code, SCA tools generally do not assert whether a vulnerable dependency makes an application unsafe (\eg because it is exploitable by an attacker) or safe (\eg because the dependency is unused).

Open source SCA tools rely on public information for their vulnerability DBs and depend on wider community efforts to update and correct this information.
Commercial tools (\eg \toolname{snyk}) may provide their own vulnerability DB, which may refine or extend the information that is available in public. Mismatch between information in vulnerability DBs is possible, often due to timing issues where one DB is updated sooner than another.  It is however in the interest of commercial vendors to eventually have their DBs aligned with public knowledge to avoid confusion among customers.
 
Like all program analyses, SCA tools suffer from precision problems, i.e. false positives. They may for instance detect dependencies to vulnerable code in a  library that is not actually reachable~\cite{mir2023effect}. This could in principle be tackled by employing more fine-grained analyses like call graph construction, although the price (in terms of computational resources needed) could be significant, and those analyses themselves have to deal with precision issues.

 
But SCA analyses are not sound either, i.e. they miss dependencies and therefore problems such as vulnerabilities associated with those dependencies~\cite{dann2021identifying}. 
 
The first source of unsoundness is late binding, i.e. applications that "discover" capabilities at runtime, leading to dependencies that are not visible in the build configurations or code SCA tools analyse.  
This is outside the scope of our study. 
 
A second cause of unsoundness is \textit{cloning}. With cloning, code is copied into the project, and these copies can carry vulnerabilities which are then hidden by the process. This can take place when an application directly clones code, or cloning is used by libraries which are then used as dependencies by downstream clients. Cloning can take place on multiple levels, from code snippets, functions, classes to entire components.  Code sharing, discussion and tutorial web sites like \textit{stack overflow}~\cite{ragkhitwetsagul2019toxic,baltes2020code} and more lately AI-based tools like \textit{Copilot} promote cloning~\cite{peng2023impact}.  With cloning, basic engineering principles like \textit{DRY (do not repeat yourself)} are violated, and in the long term the (lack of) maintenance of cloned code is highly problematic.  For vulnerability detection, a particular problem is that many clones are not perfect, i.e. they are often somehow transformed, and generally lack provenance. I.e., the original source of the clone is often opaque, and tools cannot reason about it. \textit{Copilot} is an extreme example where abstraction and aggregation is used to produce code from potentially large amounts of input sources. 
 
 However, there are also advantages to cloning, and cloning may even be used in order to make code more secure and reliable. For instance, if a dependency is only used for the purpose of using  a rather small and trivial piece of functionality from an otherwise large component, then cloning can be a sensible strategy as it may reduce the size of a product to be deployed, and may also reduce its attack surface by removing now redundant functionalities. 
 
 In the case of Java, there is an additional problem, a relative of the infamous \textit{dll hell} problem~\cite{dick2018dll}. Large dependency networks may lead to conflicts between different versions of the same class added via multiple dependency paths~\cite{wang2018dependency}. Often, the problems resulting from this only manifest at runtime when classes are loaded and linkage related errors caused by binary incompatibility occur. API changes causing this problem are common~\cite{raemaekers2014semantic,jezek2015java},  poorly understood by developers~\cite{dietrich2016java}, and therefore expensive for projects. 
 
 A common solution for this problem is \textit{shading} -- a variant of cloning where entire packages are cloned and renamed. Even the Java standard library employs shading, for instance, the OpenJDK version 16 contains shaded versions of \textit{sax} (an XML parser library) and \textit{asm} (a bytecode engineering library) in packages with names starting with \texttt{jdk.internal.org.xml.sax} and \texttt{jdk.internal.\-org.\-objectweb.asm}, respectively~\footnote{\url{https://github.com/AdoptOpenJDK/openjdk-jdk16/tree/master/src/java.base/share/classes/jdk/internal/org/}}.
 
 The Java / Maven community acknowledges this issue by providing tools like the \textit{Maven shade plugin}~\footnote{\url{https://maven.apache.org/plugins/maven-shade-plugin/}}. Here, shading is automated and performed during the build. A dependency to be shaded and the packages to be renamed are declared in the build file (\textit{pom.xml}), and therefore remain visible to SCA tools. We refer to this as \textit{build time shading} (\textit{b-shading} for short). 
 
 There are cases where other approaches to shading are used - we will provide plenty of examples later in the evaluation section. Here, shading is done by means of refactoring and code organisation tools like IDEs, we therefore refer to this as \textit{designtime shading (d-shading)}. One reason might be a misunderstanding of dependency mechanisms by engineers.  However, there might also be more sophisticated reasons to use d-shading. Tools like the shade plugin are static analysis tools, and as such can at best be expected to be \textit{soundy}~\cite{livshits2015defense}. The prevalence of dynamic language features in Java is a known challenge for static analysis tools, and leads to a considerable amount of false negatives~\cite{sui2020recall}. In particular, tools like the shade plugin have to rewrite the bytecode of the code to be shaded, and change references (supertypes, method and field descriptors, etc) to the new package names. If such references are missed due to reflective references being present, builds will fail or programs may exhibit unexpected runtime behaviour. In this case, using d-shading might be a sensible choice. 
 
 However, d-shading results in blind spots for tools that rely on declared dependencies to infer the presence of vulnerabilities. The question arises whether this is common, and in particular, whether this poses a security risk. This is the question we set out to study.  
 
 The rest of this paper is organised as follows. We start with a discussion of related work. In order to gauge how common shading is, we then report on a modest experiment on the use of the shade plugins in poms found on GitHub. In Section~\ref{sec:detection} we describe the tool pipeline we have developed in order to detect design time shading of Maven artifacts with a focus on detecting vulnerabilities in the shaded components which are missed by SCA tools.  The evaluation is split into two parts - we first describe the methodology used, and then discuss results. We discuss the disclosure procedure we followed in Section~\ref{sec:disclosure}, and finish with a short conclusion.

\section{Related Work}
\label{sec:relatedwork}

\subsection{Detecting and Managing Vulnerable Dependencies}

A study by Contrast Security investigated vulnerabilities in Java applications and found that ``custom Java applications contain from 5 to 10 security vulnerabilities per 10,000 lines of code.'' \cite{williams2014unfortunate}.  They point out that it generally has to be assumed that vulnerabilities are present in all applications, but on the other hand, that this does not always render applications unsafe. 

Mir et al~\cite{mir2023effect} point out that ``less than 1\% of packages have a reachable call path to vulnerable code in their dependencies'', alerting to precision problems of dependency-based SCA. However, those results have to be interpreted with caution. The underlying call graph analysis is based on Opal~\cite{eichberg2014software}, configured to run the rather inaccurate (but fast) class hierarchy analysis (CHA, \cite{grove1997call}). This is likely to miss many dynamic call graph edges~\cite{sui2020recall} which are exploited in vulnerabilities. As an  example, consider CVE-2015-6420. This vulnerability can be exploited by deserializing objects from an incoming stream, and therefore the call graph path from application classes to vulnerable classes in this library is highly obfuscated, and unlikely to be detected by CHA-based (or any other scalable) call graph construction method. The work by Wu et al~\cite{wuunderstanding} is related, with similar limitations.

Kula et al~\cite{kula2018developers} studied how developers respond to vulnerabilities being detected in dependencies they rely on. They found that most of the time outdated dependencies are kept, and developers are unlikely to respond to security advisories~\cite{kula2018developers}. Similar results, reporting significant delays to upgrade vulnerable dependencies, were also observed for other ecosystems, for instance by Decan et al for NPM~\cite{decan2018impact} and Alfadel et al for Python~\cite{alfadel2023empirical}. 

Mirhosseini and Parnin studied whether automated pull requests (PRs) are effective to speed up upgrades~\cite{mirhosseini2017can}. This mechanism is often deployed by composition analysis tools like \textit{dependabot}. In general, they found that PRs do speed up upgrades, although the merge rate is still surprisingly low at around a third of all PRs.  Alfadel et al studied particular PRs made by a popular SCA tool, GitHub's \textit{dependabot}, and found a significantly higher acceptance (merge) rate of about two thirds. This study considered only NPM projects.

Dann et al \cite{dann2021identifying} studied several OSS vulnerability scanners (\textit{OWASP dependency check, Eclipse steady, snyk, black duck, WhiteSource}) and evaluated their performance on a set of 7,024 projects collected by SAP. They found limitations of the tools to deal with several modifications (re-compilation, re-bundling, metadata-removal and re-packaging) of the original vulnerable projects. Their observations are consistent with ours, and the respective  modifications roughly correspond to our notions of cloning and shading.  

Bui et al developed \textit{vul4j} \cite{bui2022vul4j}, a dataset consisting of 79 reproducible vulnerabilities from 51 open-source projects. Reproducibility is achieved via proof-of-vulnerability (POV) tests. This is the same approach we are using to confirm the presence of a vulnerability. We use some suitable parts of this dataset for our evaluation, details will be discussed in Section~\ref{sec:evaluationmethodology:dataset}.

Ponta et al \cite{ponta2018beyond} propose a hybrid code-centric vulnerability detection that overcomes the limitations (here mainly seen as the low precision)  of meta-data based SCA approaches. Their analysis uses code changes introduced by security fixes.  The tool resulting from this is \textit{vulas}, later renamed to \textit{steady}.  We used \textit{steady} in the evaluation (Section \ref{sec:evaluationmethodology:scatoolselection}) and our results suggest that it is complementary to our approach. 

Our approach depends on the existence of proof-of-vulnerability (POV) projects, and would therefore  benefit from the automated generation of exploits. Initial work in this area, based on test case generation using genetic algorithms, has been proposed by Iannone et al ~\cite{iannone2021toward}.

\subsection{Clone Detection}
Research into code clone detection has established a classification for levels of clone similarity: type-1 clones are identical except for layout (whitespace) and comments; type-2 clones are syntactically equivalent, allowing for renaming of variables, functions, types, etc.; type-3 clones are syntactically similar, additionally allowing for some statements to be added or removed.

Clone detection is used to improve or enforce software development quality standards by detecting unwanted copies of code leading to maintainability or licensing issues, and, in academic settings to detect plagiarism. There is a vast amount of research in this field, covered in surveys such as  \cite{roy2009comparison,rattan2013clones}. 

We use type-2 clone detection as a proxy to detect compositional clones, \ie the practice of copying (parts of) existing software libraries into projects.

Binary code similarity~\cite{Haq2021} can be seen as an instance of clone detection and related to our work. Comparing code at the binary (or bytecode) level comes with the challenge of variations introduced by different (versions of) compilers, different compile-time transformations, and different compile environments.

In particular for Java, clone detection in bytecode has been studied by Dann et al~\cite{dann2019sootdiff}. They address the problem by translating bytecode into an intermediate, soot-based format that can abstract from the particularities of different compilers to some extent. We did consider using a similar approach, however, as source code is readily available in Maven, a traditional AST-based clone detection appears to be the better choice as those problems can be avoided.   

Closely related to it, and also targeting the Java open-source ecosystem is SCA-related research focusing on libraries included in released Android applications under the term \emph{third-party library detection}~\cite{Zhan2021}. Note that in this context, research tries to solve the harder problem of creating an analysis that is resilient to hiding and obfuscation of libraries. It does this using similarity search techniques based on features (\eg class dependency structure, method signatures, control-flow graphs) extracted from bytecode.

\section{Prevalence of Buildtime Shading}
\label{sec:prevalence}

To gauge how widespread the practice of shading is, we first focus on b-shading, shading performed at build time. If Maven is used as a build system, this can be achieved by using the Maven shade plugin. 
To study the prevalence of b-shading, we used a dataset consisting of Maven build files (poms, i.e., \texttt{pom.xml}) collected as follows: (1) As a starting point, the \textit{library.io} dataset~\footnote{\url{https://libraries.io/}} released on 12 January 2020 was used. (2) The projects were then filtered for projects using Maven / Java, and having a GitHub repository. (3) \texttt{pom.xml} files were extracted from the respective repositories.  This produced an initial  set of 103,358 poms. We selected poms using the shade plugin with the following XPath query: 

\begin{scriptsize}
\vspace{0.2cm}\noindent
\verb|//plugin/artifactId[text() = 'maven-shade-plugin']|
\vspace{0.2cm}
\end{scriptsize}

This query yields 3,693 poms (3.57\%).  We then analysed how often classes are relocated into different packages, by querying those poms with the following XPath query: 
 
\vspace{0.2cm}\noindent
\begin{scriptsize}
\begin{verbatim}
//plugin/artifactId[text() = 'maven-shade-plugin']/parent::node()//relocations
\end{verbatim}
\end{scriptsize}

\vspace{0.2cm}

We found 808 poms (0.78\% of all poms) using relocations.  The results indicate that build time shading is commonly used, supporting the claim that there are valid use cases for shading in general, and for package renaming in particular.

This raises the questions of how common shading is that does not use plugins (and therefore does not state the dependency), how it can be detected, and what the security implications of this are.

\section{Blindspot Detection}
\label{sec:detection}

\subsection{Overview}
\label{ssection:detection:overview}

We describe the processing pipeline we have developed and used to detect clones and shaded artifacts with known vulnerabilities here. It takes an artifact and a vulnerability as input, and produces a list of artifacts and projects demonstrating the presence of the provided vulnerability in those artifacts.  The focus of the tool design is on precision (avoiding false positives)\footnote{Precision here is defined with respect to the presence of vulnerable code, not taking into account whether it is actually exploitable in the context of a particular application. I.e. we want to find clones which are \textit{as vulnerable} as the original component for the respective CVE.}, and being lightweight. In particular, it does not require the acquisition, construction and maintenance of a separate index. Instead, it can work with an existing index as long as it makes the information required (source code, poms, artifacts searchable by class names) available through an API. Our aim is to demonstrate that with some fairly simple tooling that goes beyond the metadata-centric approach used by most SCA tools, more vulnerable artifacts can be detected. We do not aim at detecting all those artifacts, and as with all program analyses, precision, recall and performance have to be balanced, as perfect non-trivial analyses are not feasible~\cite{rice1953classes,ernst2003static}. 


The key ideas our tooling is based on are:
\setdefaultleftmargin{0cm}{1cm}{1cm}{1cm}{1cm}{1cm}

\begin{enumerate}
	\item We do not depend on indexing the Maven repository, instead, we work with the existing repository via the Maven Central REST API~\footnote{\url{https://central.sonatype.org/search/rest-api-guide/}}.  
	\item We extract a signature of components to be used to identify potential clones using a lightweight method based on unqualified, characteristic class names. This can then be directly used in repository queries.
	\item We use a custom AST-based clone detection to identify clones, including those that may have repackaged classes. 
	\item We  use tests to verify the presence of vulnerabilities in clones, and automate the adaptation of those tests for clones, and the evaluation of test results.  This leads to a high precision of the vulnerability detection.
\end{enumerate}

\subsection{Inputs}

Our analysis requires the following inputs: 

\begin{enumerate}
	\item An artifact $art_0$ identified by its group-artifact-version (GAV) coordinates $gav_0$ within the Maven repository
	\item A known vulnerability $vul$ identified by a CVE  
	\item A proof-of-vulnerability (POV) Maven project $pov$ that has a direct dependency on $art_0$ and one or many tests demonstrating the presence of $vul$. Those tests demonstrate the presence of the respective vulnerability, their evaluation leads to an expected test \textit{signal} (usually \textit{success} or \textit{failure}). 
\end{enumerate}

The project $pov$ is optional in the sense that the tool chain can be run without it. Its purpose is to make the analysis results precise.

\subsection{Pipeline}



Our analysis pipeline consists of the following steps, fetch steps imply interaction with the Maven REST API:

\begin{enumerate}
	\item \textbf{Fetch binaries} -- Fetch the binary (jar) $art_0.bin$  of $art_0$.
	\item \label{pipeline:fetchsources}  \textbf{Fetch sources }-- Fetch the source code  $art_0.src$  of $art_0$.
	\item \label{pipeline:extractclasses} \textbf{Select classes} --  Extract a set of classes $cl.query$  from the  $art_0.bin$ and/or $cl.vul$ to be used in queries. Those are non-qualified class names (i.e., package names are omitted). 
	\item \textbf{Fetch class matches} -- For each class in $cl \in cl.query$,  fetch a set of artifact coordinates (GAVs) $match_{cl}$  of artifacts containing a class with this (unqualified) name. 
	\item \label{pipeline:classconsolidation} \textbf{Consolidate matches} -- Consolidate all sets $match_{cl}$ into a single set $match$ .
	\item  \label{pipeline:query}  \textbf{Fetch \textit{match} poms} -- For each artifact $art  \in match$, fetch the pom $art.pom$ . 
	\item  \label{pipeline:prune}  \textbf{Remove dependents of original artifact} -- For each artifact $art  \in match$, analyse the pom $art.pom$, and  if it contains reference to $art_0$, remove it.  
	\item \textbf{Fetch \textit{match} sources} --  For each artifact $art  \in match$, fetch the source code $art.src$. 
	\item \textbf{Run clone analysis} -- For each artifact $art  \in match$, run a clone analysis comparing $art_0.src$ and $art.src$, and if the result is negative, remove $art$ from $match$
	\item \textbf{Instantiate POV} -- For each artifact $art  \in match$, instantiate   $pov$  by cloning the project and replacing the dependency to  $art_0$ by a dependency to $art$, resulting in a project $pov(art)$.
	\item \textbf{Verifying the Vulnerability} -- For each artifact $art  \in match$, run \texttt{mvn test} on $pov(art)$.  If this succeeds, the presence of the vulnerability is confirmed, and $art$ is added to the result.
	
\end{enumerate}

We describe the more interesting steps  briefly in the rest of this section, and state the settings we used in the evaluation section for the steps that are configurable. We do not claim that those settings are optimal, but that they produce a reasonable yield in terms of artifacts with vulnerabilities discovered with modest computational resources.


\subsection {Class Selection}

We use unqualified class names as fingerprints to identify potential clones. There are two reasons for this: (1) the Maven REST API supports queries by unqualified class names (2) unqualified class names are not changed when relocating code during shading. 

When working with a remote index, using all classes is not a good strategy as each class name is then used in a query, as each class may result in multiple network calls. We have used a simple approach to look for signature classes with names likely to be unique. For instance, a short name like \texttt{Utils} is likely to be used by many components. However, something like \texttt{JSONDriverManagerFactory} (hypothetical) is more likely to be unique. The heuristic used is to count camel case tokens in class names, and look for classes with a high count. In the example above, the count for \texttt{JSONDriverMan\-agerFactory} is 4, whereas the count for \texttt{Utils} is 1.

The default strategy we have employed is to sort class names by token length, and to use the top 10 class names. 


\subsection{Fetch Class Matches}
\label{ssec:detection:query}

For each class name identified in step \ref{pipeline:extractclasses}, an API query is used to fetch artifacts containing one or more classes with this name. The API uses paging, and limits the number of results returned by each query to 200. We use 5 pages of 200 results each, i.e. a maximum of 1,000 artifacts per class is analysed.

\subsection{Query Consolidation }

The process described above results in 10 query result sets with up to 1,000 artifacts in each. A consolidation strategy identifies the artifacts likely to represent clones. Strategies like intersection or union of result sets are possible, the union is likely to contain many accidental matches that contain only a single matching class. The other extreme, the intersection, may exclude many artifacts that only partially clone the original artifact, but could still contain all classes necessary to exploit a vulnerability. The strategy we have used is that an artifact must occur in at least two result sets, i.e. it must contain at least two classes with names matching classes in the original artifact selected for querying.  

\subsection{Remove Dependents of Original Artifact}

This step is performed in order to remove artifacts that declare a dependency to the original artifact. Those are less interesting and may even be considered as effective false positives by engineers~\cite{sadowski2018lessons} as SCA tools usually detect vulnerabilities propagated through such dependencies. For this purpose we acquire and analyse the pom of the artifact. The pom analysis is looking for three patterns: (1) There is no reference in the dependency section to the original artifact. (2) There is no reference to the original artifact within the shade plugin.
(3)The group id and artifact id of the clone candidate are different from the group and artifact ids of the original artifact.

The last rule ensures that the tool does not produce results representing different versions of the original artifact.  Our analysis also includes references in parent poms for artifacts generated by multi-module projects.

\subsection{Clone Analysis}

The clone analysis used is AST-based. I.e., candidates classes are parsed and the two ASTs are simultaneously traversed. Our method is a type-2 clone detection~\cite{roy2009comparison}, i.e., we are looking for isomorphic structures but allow some variations in types and comments.

Nodes corresponding to comments are ignored as authors may change comments (for instance, to alter copyright or authorship notices, or to add comments about the origin of the code). For nodes corresponding to type names, the scopes (package names) are ignored.

\subsection{Instantiating the POV Project}
\label{ssect:detection:pov}

For each artifact $art  \in match$, we instantiate the POV project $pov$ by cloning it, replacing the dependency in the pom to  $art_0$ by a dependency to $art$, and replacing references to fully qualified class names in classes defined in $pov$ (in particular, tests) if classes are re-packaged by the clone.  The mapping of classes from old to new packages, representing their relocation, is provided by the clone analysis and is used here to update class references, \eg in import statements.

Consider for instance the test used to demonstrate the presence of  CVE-2022-38751, a DOS vulnerability in \textit{snakeyaml}, shown in Listing \ref{test:snakeyaml}~\footnote{The code listings are shortened for brevity}. The structure of the test is straight-forward -- parse a malicious payload (\textit{CVE-2022-38751.yml}), and check that this leads to a stack overflow error. If this leads to some other error or exception (such as an \textit{IllegalArgumentException}), the test fails, indicating that the vulnerability is not present. 

\begin{lstlisting}[label={test:snakeyaml},caption={Testing CVE-2022-38751 (snakeyaml)}, language={Java}]
import org.yaml.snakeyaml.Yaml;
// more imports ommitted 
public class ConfirmVulnerabilitiesTests {
    @Test public void confirmCVE202238751 () {
        assertThrows(
            StackOverflowError.class,
            () -> parse("CVE-2022-38751.yml")
        );
    }
    static void parse (String input) throws IOException {
        FileReader reader = new FileReader(new File(input));
        new Yaml().compose(reader);
    }
}
\end{lstlisting}

Also note the \texttt{import} statement in line 1. If we find a clone, the original test can be copied, and instantiating the POV project is merely a matter of replacing the dependency on \textit{snakeyaml} by a dependency on the respective clone.  If during the clone detection phase clones are detected in different packages, then the import statement needs to be changed as well. This is done by manipulating the ASTs of the respective source files. 

The signal for this particular POV test is \textit{success}, i.e. the test succeeding indicates that the vulnerability is present. This is often an intuitive approach to write POV tests as the effect of the vulnerability is used directly as a test oracle. However, often POV projects can be sourced from vulnerability patches. In this case, tests are often designed as regression tests, failing to indicate the vulnerability is present and a patch is required. In this case, the expected signal is \textit{failure}.  To support this test signal abstraction, the test signal is documented in POV projects. 

A list of POV projects used in our validation can be found here: \url{https://github.com/jensdietrich/xshady}.

\subsection{Verifying The Presence of a Vulnerability}

For each artifact, $art  \in match$, we run \texttt{mvn test} on $pov(art)$.  If the test signals recorded are identical to the signals of the original POV project, the presence of the vulnerability is confirmed.  This is done by analysing generated \textit{surefire} reports.  Since the success of \texttt{mvn compile} is a prerequisite for testing, we run this phase first to filter out projects that cannot be build first, avoiding the more expensive test runs. This step has the sole purpose of optimising pipeline performance.

A particular issue that needs to be taken into account is that tests may result in four states -- \textit{success}, \textit{failure}, \textit{error} and \textit{skip}. Builds with tests succeed if all tests are in a \textit{success} or \textit{skip} state. This is an optimistic ``did not fail assumption''. However, we found that it is often practical or even necessary to use assumptions in tests confirming vulnerabilities. 

For instance, consider the test confirming CVE-2022-25845 in \textit{fastjson}, shown in Listing~\ref{test:fastjson}. The test confirms the execution of an OS command triggered by parsing a document, the command used here is ``\texttt{touch foo}''. This command is defined in the JSON document to be parsed (CVE-2022-25845.json), the \textit{@BeforeEach} fixture is used to erase the file if present. This OS command is only available on unix-like operating systems, and the vulnerability can only be exploited for certain JRE versions. This is encoded using JUnit precondition (assumption) annotations (lines 10-11), and tests are skipped (instead of failed) if those conditions are not satisfied. Therefore, the analysis needs to confirm that all tests have succeeded, which is a stricter requirement (i.e. stricter than the default \textit{surefire} behaviour).


\begin{lstlisting}[label={test:fastjson},caption={Testing CVE-2022-25845 (fastjson)}, language={Java}]
import com.alibaba.fastjson.JSON;
// more imports ommitted 
public class ConfirmVulnerabilitiesTests {
    @BeforeEach public void clearGeneratedFile() {
        File file = new File("foo");
        if (file.exists()) {
            Assumptions.assumeTrue(file.delete());
        }
    }
    @Test @EnabledOnOs({OS.MAC,OS.LINUX})
    @EnabledForJreRange(min=JRE.JAVA_8,max=JRE.JAVA_11)
    public void confirmCVE202225845 () throws Exception {
        Path generatedFile = Path.of("foo");
        Assumptions.assumeFalse(Files.exists(generatedFile));
        Path payload = Path.of("CVE-2022-25845.json");
        Assumptions.assumeTrue(Files.exists(payload));
        String json = Files.readString(payload);
        JSON.parse(json); 
        Thread.sleep(1000);
        assertTrue(Files.exists(generatedFile));
    }
}
\end{lstlisting}

\section{Evaluation Methodology}
\label{sec:evaluationmethodology}

\subsection{Dataset}
\label{sec:evaluationmethodology:dataset}

The selection of a set of CVEs used for evaluation was driven by the following considerations: (1) to select widely used artifacts, as determined by the number of downstream clients reported by Maven (2) to select CVEs of different types, namely vulnerabilities exploitable for remote code execution (RCE) and denial of service (DOS) attacks (3) to include some high-impact vulnerabilities that have been exploited in the wild such as \textit{log4shell} (4) to select libraries from different domains (5) to select CVEs in libraries we considered as good candidates for cloning. We argue that single-purpose libraries that do not have significant further upstream dependencies and do not use dynamic programming features are better candidates for cloning as cloning as this caps the complexity of the process of integration. In particular, we expect that complex application frameworks such as \textit{spring} and \textit{struts} are not good candidates for cloning. Since our aim was to make CVEs testable in order to design a precise analysis, we furthermore gave preference to CVEs with available proof-of-vulnerability projects we could then reuse (usually with some modifications). In particular for vulnerabilities that have a high severity, such projects often exist. Sometimes projects covering entire classes of vulnerabilities can be used for this purpose, a good example is \textit{ysoserial}~\footnote{\url{https://github.com/frohoff/ysoserial}} that also  contains a POV for CVE-2015-6420 which we used in a slightly modified, testable form. 

We selected 9 CVEs manually to fit those criteria. We then complemented this dataset with CVEs from \textit{vul4j}~\cite{bui2022vul4j}, an independent dataset consisting of vulnerabilities, artifacts and vulnerability patches including regression tests.  \textit{Vul4j} consists of 79 CVEs.  We found that most are not suitable for our purpose for different reasons: many CVEs in \textit{vul4j} are related to application frameworks, 27 alone are from 3 frameworks (\textit{spring}, \textit{struts} and \textit{jenkins}).  Some do not have tests (e.g. CVE-2016-3720 and CVE-2017-5662), the vulnerability cannot be reproduced with the provided test(s) (e.g., CVE-2019-10173, CVE-2018-1000850) or the component flagged as vulnerable is not in Maven central (e.g., CVE-2018-17202, CVE-2018-17201).  In the end we added 20 additional CVEs from \textit{vul4j}. The total dataset of 29 is depicted in Table~\ref{tab:dataset}. The table shows the wide coverage of our dataset with respect to vulnerability types, years when the CVE was assigned, and vulnerable components. 


\begin{table*}[]
	\begin{tabular}{|p{1.5cm}|p{6.2cm}p{3.8cm}p{3.8cm}|}
		\hline
		severity & RCE, XSS   & DOS & Other  \\         
		\hline                                                                 
		critical & 
		\textit{2022-25845 (fastjson)}, \textit{2022-42889 (c-text)}, \textit{2021-44228 (log4j)}, \textit{2015-6420 (c-collections)}, 2020-1953 (c-config), 2017-18349 (fastjson), 2016-0779 (tomee), 2015-7501 (c-collections) &  &    2016-6798 (sling)                                                                                          \\
		high     & \textit{2016-2510 (beanshell) }                                                                                                                                       &  \textit{2022-45688 (json.org)}   &       2019-12402 (c-compress), 2019-0225 (jspwiki), 2016-6802 (shiro), 2016-7051 (jackson)                                                           \\
		medium   &   2017-15717 (sling), 2016-5394 (sling), 2015-6748 (jsoup)    & \textit{2022-38749 (snakeyaml)} , \textit{2022-38751 (snakeyaml)}, \textit{2018-10237 (guava)}, 2018-11771 (c-\-com\-press), 2018-1324 (c-compress), 2018-8017 (tika)  &  2021-29425 (c-io), 2018-1002201  (zt-zip) \\
		n/a  &  & 2014-0050 (c-fileupload) & 2013-2186 (c-fileupload), 2013-5960 (esapi) \\  \hline
		
	\end{tabular}
	\caption{\label{tab:dataset}Dataset used in the evaluation, organised by severity (NVD base score as of 20 Sept 2023) and type.\textit{ c-} stands for \textit{``apache commons''}. Vulnerabilities not sourced from \textit{vul4j} are highlighted italic. CVE- prefixes are omitted for brevity.}
\end{table*}

\subsection{SCA Tool Selection}
\label{sec:evaluationmethodology:scatoolselection}

There are numerous tools available to detect the presence of vulnerable dependencies in software projects. During evaluation we used a curated set of SCA tools to do the following: 

\begin{enumerate}
\item To confirm that the tool(s) can detect the vulnerability in the original artifacts.
\item To confirm that some / all tools fail to detect the vulnerability in some / all clones. 
\end{enumerate}

The SCA tools used are listed in Table~\ref{tab:scatools}. We selected them in order to provide a variety of detection implementations, while aiming to increase the coverage of vulnerability DBs and keeping the effort of running multiple tools manageable. Some tools have the option of either invoking them from the command line (cli) or integrating scanning with the build process (plugin), thus we perform evaluation with tools in both categories.
We expect both, the functionality of these SCA tools, and contents of their DBs to overlap, but not to be equivalent either. Reasons for this are discussed in Section~\ref{sec:introduction}.
As an example, adding GitHub's \toolname{dependabot} would not have increased DB coverage of our evaluation because its vulnerability DB, GHSA, is already covered by our selection.

\begin{center}
	\begin{table}
		\begin{tabular}{|lll|}
			\hline
			tool & mode & databases (java) \\
			\hline
			OWASP Dependency Check (owasp) & plugin & NVD, OSS Index \\
			snyk & cli & proprietary \\
			grype & cli & NVD, GHSA \\
			Eclipse Steady (steady) & plugin & Project KB \\
			\hline
		\end{tabular}
		\caption{SCA Tools used}
		\label{tab:scatools}
	\end{table}
\end{center}



\section{Evaluation Results}
\label{sec:evaluationresults}

\subsection{Pipeline Performance}

As described earlier in Section \ref{ssec:detection:query}, we start with fetching 1,000 potentially matching artifacts for each class name,  up to 10,000 artifacts in total (for 10 classes). We record the number of artifacts after each step of processing and filtering. We report a summary of the results in Table~\ref{tab:pipelinestats}, both for artifacts, and for artifacts aggregated by ignoring versions.  This indicates that we find vulnerable clones for 18 / 29 components (column 10). Interestingly, there is one vulnerability where we do not find any matching artifact using any of the initial queries.  This is for CVE-2016-6802 , a vulnerability in \textit{org.apache.shiro:shiro-all:1.3.1}. This artifact does not define classes itself, but bundles classes from other \textit{shiro} modules. Our algorithm is currently not able to extract classes to be used as queries here.  The last column shows the number of vulnerable clones where shading was applied and  packages have been renamed.

\begin{figure*}
	\includegraphics[width=\textwidth]{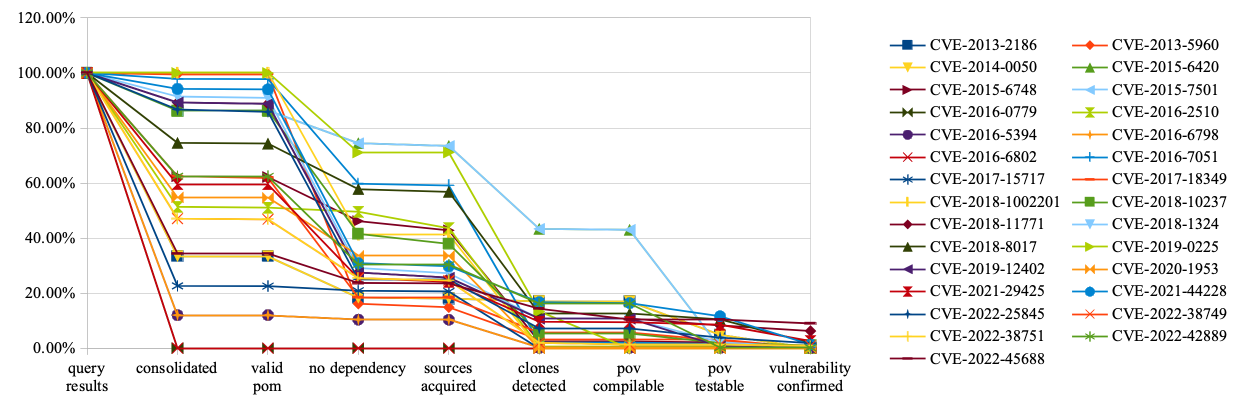}
	\caption{Pipeline throughput relative to the number of artifacts initially fetched}
	\label{fig:pipeline}
\end{figure*}

Figure~\ref{fig:pipeline} depicts the pipeline throughput for each stage of processing -- scaled to 100\% -- the initial set of artifacts acquired via the Maven API.  For almost all artifacts consolidated (i.e. being the the result set for more than one queries) poms can be acquired. At the no dependency stage, artifacts with dependency on the original vulnerable artifacts are filtered out as they are likely to represent effective false positives. Our mechanism to acquire sources generally works well, but there are cases where sources cannot be located using the REST API, and may just be missing. We also observed a very few cases where we were able to acquire sources, but not able to unpack the respective (potentially corrupted) archives. In the actual clone detection step, a significant number of artifacts is removed. Note that that this is a simple, very fast static analysis. The last stages to establish whether the instantiated POV project can be built and tested successfully (i.e., the POV signal being confirmed) is a significantly more expensive analysis as this requires a build and multiple interactions with the Maven repository to resolve and fetch the dependencies of the instantiated POV project.

\begin{table*}[]
	\begin{tabular}{|lllllllllll|} \hline
		\textbf{} & \textbf{\begin{tabular}[c]{@{}l@{}}query\\ results\end{tabular}} & \textbf{\begin{tabular}[c]{@{}l@{}}consol-\\ idated\end{tabular}} & \textbf{\begin{tabular}[c]{@{}l@{}}valid\\ pom\end{tabular}} & \textbf{\begin{tabular}[c]{@{}l@{}}no depend-\\ ency\end{tabular}} & \textbf{\begin{tabular}[c]{@{}l@{}}sources\\ acquired\end{tabular}} & \textbf{\begin{tabular}[c]{@{}l@{}}clones\\ detected\end{tabular}} & \textbf{\begin{tabular}[c]{@{}l@{}}pov\\ compilable\end{tabular}} & \textbf{\begin{tabular}[c]{@{}l@{}}pov\\ testable\end{tabular}} & \textbf{\begin{tabular}[c]{@{}l@{}}vulnerability\\ confirmed\end{tabular}} & \textbf{\begin{tabular}[c]{@{}l@{}}shaded\end{tabular}} \\ \hline
			min             & 0        & 0        & 0        & 0      & 0      & 0      & 0      & 0     & 0     & 0     \\
			max             & 4,675    & 2,666    & 2,664    & 1,367  & 1,267  & 669    & 574    & 483   & 419   & 190   \\
			\textgreater{}0 & 28       & 27       & 27       & 27     & 27     & 24     & 23     & 23    & 18    & 8     \\
			avg             & 1,840.34 & 1,029.10 & 1,025.83 & 548.14 & 527.97 & 171.72 & 163.07 & 61.07 & 25.07 & 12.72 \\
			sum             & 53,370   & 29,844   & 29,749   & 15,896 & 15,311 & 4,980  & 4,729  & 1,771 & 727   & 369   \\ \hline                                                   
		\multicolumn{11}{|c|}{versions ignored} \\
		\hline
			max             & 410    & 294   & 293   & 171   & 165   & 44    & 34    & 26   & 26   & 13   \\
			avg             & 153.86 & 81.62 & 81.03 & 53.24 & 50.41 & 11.10 & 10.17 & 6.38 & 2.97 & 1.00 \\
			sum             & 4,462  & 2,367 & 2,350 & 1,544 & 1,462 & 322   & 295   & 185  & 86   & 29    \\ \hline                                                                    
	\end{tabular}
	\caption{Pipeline throughput statistics -- artifacts (GAV)  in top half and components (GA -- versions aggregated) in lower half, counts after each stage of processing}
	\label{tab:pipelinestats}
\end{table*}

\subsection{ Vulnerable Artifacts Found}

We find vulnerable clones for 18 of the 29 CVEs studied, details are shown in Table~\ref{tab:pipelinestats}.  The total number of vulnerable artifacts found is 727 across all CVEs. Often those are different versions of the same artifact, after deduplicating those and ignoring versions (\textit{``aggregated''}), the number drops to 86. The highest number of vulnerable components is detected for CVE-2022-45688, a DOS vulnerability in \textit{org.json:json:20230227}, with 419 vulnerable clones detected (26 aggregated).  JSON parsing and encoding is a popular requirement for data persistence and exchange, and the compact, self-contained nature of \textit{json.org} make it a good candidate for cloning. 

The last column in Table~\ref{tab:pipelinestats} reports the vulnerable clones where shading with renaming of packages has been applied. This is the case for 45.35\% of the detected artifacts. This number is high and particularly significant as the embedded code is now less likely to be spotted by developers or tools. 


For the infamous CVE-2021-44228  log4j vulnerability we detected  15 clones (3 when versions are ignored), none of them using shading with package renaming.

%
%
%

Note that some artifact names are masked due to the disclosure process we follow, this will be described in detail in Section~\ref{sec:disclosure}.

\subsection{SCA Results}
\label{ssec:evaluationresults:sca}

We have set out to investigate whether existing SCA tools and analyses find vulnerabilities in clones. To set a baseline, we first had to established whether the selected SCA tools (see Section \ref{sec:evaluationmethodology:scatoolselection}) can detect the vulnerability in the original artifact~\footnote{Often, vulnerabilities are reported to entire version ranges. The artifact we consider in this case is the latest within the range. The precise coordinates can be found in the POV repository (\url{https://github.com/jensdietrich/xshady}), in the \texttt{pom.xml} dependency settings in POV project for the respective CVE.}. The respective SCA tool reports can be found in the POV repository (\url{https://github.com/jensdietrich/xshady}), Table~\ref{tab:sca:originals} summarises the results from those reports. 

This shows that \textit{snyk}, \textit{owasp} and \textit{grype} can detect most vulnerabilities, with minor variations between them. \textit{Steady} performs worse, likely caused by the fact that it uses a separate knowledge base that is incomplete and not up-to-date. 

\begin{table}[]
	\begin{tabular}{|lllllll|}
		 \hline
		cve              & steady & snyk & owasp & grype & any & all \\ \hline
		CVE-2013-2186    & $\times$      & $\checkmark$    & $\checkmark$     & $\checkmark $    & $\checkmark $  & $\times$   \\
		CVE-2013-5960    & $\times$      & $\checkmark $   & $\times$     & $\checkmark $    & $\checkmark $  & $\times$   \\
		CVE-2014-0050    & $\times$      & $\checkmark $   & $\checkmark $    & $\checkmark $    & $\checkmark $  & $\times$   \\
		CVE-2015-6420    & $\times$      & $\checkmark $   & $\checkmark $    & $\checkmark $    & $\checkmark $  & $\times$   \\
		CVE-2015-6748    & $\checkmark $     & $\checkmark $   & $\checkmark $    & $\checkmark $    & $\checkmark $  & $\checkmark $  \\
		CVE-2015-7501    & $\times$      & $\checkmark $   & $\times$     & $\checkmark $    & $\checkmark $  & $\times$   \\
		CVE-2016-0779    & $\checkmark $     & $\times$    & $\checkmark $    & $\times$     & $\checkmark $  & $\times$   \\
		CVE-2016-2510    & $\checkmark $     & $\checkmark $   & $\checkmark $    & $\checkmark $    & $\checkmark $  & $\checkmark $  \\
		CVE-2016-5394    & $\times$      & $\checkmark $   & $\checkmark $    & $\checkmark $    & $\checkmark $  & $\times$   \\
		CVE-2016-6798    & $\times$      & $\times$    & $\times$     & $\checkmark $    & $\checkmark $  & $\times$   \\
		CVE-2016-6802    & $\checkmark $     & $\checkmark $   & $\checkmark $    & $\checkmark $    & $\checkmark $  & $\checkmark $  \\
		CVE-2016-7051    & $\checkmark $     & $\checkmark $   & $\checkmark $    & $\checkmark $    & $\checkmark $  & $\checkmark $  \\
		CVE-2017-15717   & $\times$      & $\checkmark $   & $\checkmark $    & $\checkmark $    & $\checkmark $  & $\times$   \\
		CVE-2017-18349   & $\checkmark $     & $\checkmark $   & $\checkmark $    & $\checkmark $    & $\checkmark $  & $\checkmark $  \\
		CVE-2018-1002201 & $\checkmark $     & $\checkmark $   & $\checkmark $    & $\checkmark $    & $\checkmark $  & $\checkmark $  \\
		CVE-2018-10237   & $\checkmark $     & $\checkmark $   & $\checkmark $    & $\checkmark $    & $\checkmark $  & $\checkmark $  \\
		CVE-2018-11771   & $\checkmark $     & $\checkmark $   & $\checkmark $    & $\checkmark $    & $\checkmark $  & $\checkmark $  \\
		CVE-2018-1324    & $\checkmark $     & $\checkmark $   & $\checkmark $    & $\checkmark $    & $\checkmark $  & $\checkmark $  \\
		CVE-2018-8017    & $\checkmark $     & $\times$    & $\checkmark $    & $\checkmark $    & $\checkmark $  & $\times$   \\
		CVE-2019-0225    & $\checkmark $     & $\times$    & $\checkmark $    & $\times$     & $\checkmark $  & $\times$   \\
		CVE-2019-12402   & $\checkmark $     & $\checkmark $   & $\checkmark $    & $\checkmark $    & $\checkmark $  & $\checkmark $  \\
		CVE-2020-1953    & $\checkmark $     & $\checkmark $   & $\checkmark $    & $\checkmark $    & $\checkmark $  & $\checkmark $  \\
		CVE-2021-29425   & $\checkmark $     & $\checkmark $   & $\checkmark $    & $\checkmark $    & $\checkmark $  & $\checkmark $  \\
		CVE-2021-44228   & $\checkmark $     & $\checkmark $   & $\checkmark $    & $\checkmark $    & $\checkmark $  & $\checkmark $  \\
		CVE-2022-25845   & $\times$      & $\checkmark $   & $\checkmark $    & $\checkmark $    & $\checkmark $  & $\times$   \\
		CVE-2022-38749   & $\times$      & $\checkmark $   & $\checkmark $    & $\checkmark $    & $\checkmark $  & $\times$   \\
		CVE-2022-38751   & $\times$      & $\checkmark $   & $\checkmark $    & $\checkmark $    & $\checkmark $  & $\times$   \\
		CVE-2022-42889   & $\times$      & $\checkmark $   & $\checkmark $    & $\checkmark $    & $\checkmark $  & $\times$   \\
		CVE-2022-45688   & $\times$      & $\checkmark $   & $\checkmark $    & $\checkmark $    & $\checkmark$   & $\times$   \\  \hline
		sum              & 16     & 25   & 26    & 27    & 29  &    \\  \hline
	\end{tabular}
    \caption{CVEs detected by various SCA tools in the original artifact associated with the CVE}
    \label{tab:sca:originals}
\end{table}

We then used the same tools to check the clones detected by our analysis. The results are shown in Table~\ref{tab:sca:clones}. Since we detect vulnerable clones in only 18/29 
vulnerabilities, this table has only 18 rows. The \textit{any} column indicates the number of clones that are detected by all four tools. This is the number of vulnerabilities projects using any of those four tools would already be able to detect. This is only 20.50\%  (149/727). However, this is still a very conservative estimate, 
and the effective detection rate from a practical point of view is much lower as projects would typically only use one, not multiple or even all of those SCA tools.  

For 8 CVEs, there is at least one clone \textit{none} of the standard SCA  tools detects.  

\begin{table}[]
	\begin{tabular}{|lllllll|}
		\hline
		CVE            & total & grype & owasp & snyk & steady & any \\ \hline
		CVE-2015-6420  & 3     & 0     & 0     & 0    & 0      & 0   \\
		CVE-2015-7501  & 3     & 0     & 0     & 0    & 0      & 0   \\
		CVE-2016-2510  & 3     & 0     & 1     & 1    & 3      & 3   \\
		CVE-2016-5394  & 1     & 1     & 1     & 0    & 0      & 1   \\
		CVE-2016-6798  & 1     & 1     & 0     & 0    & 0      & 1   \\
		CVE-2016-7051  & 7     & 0     & 7     & 0    & 1      & 7   \\
		CVE-2018-10237 & 31    & 8     & 21    & 0    & 17     & 26  \\
		CVE-2018-11771 & 92    & 0     & 3     & 0    & 1      & 3   \\
		CVE-2018-1324  & 2     & 0     & 2     & 0    & 0      & 2   \\
		CVE-2018-8017  & 17    & 0     & 0     & 0    & 3      & 3   \\
		CVE-2019-12402 & 1     & 0     & 1     & 0    & 0      & 1   \\
		CVE-2021-29425 & 56    & 0     & 5     & 1    & 1      & 5   \\
		CVE-2021-44228 & 15    & 0     & 14    & 0    & 15     & 15  \\
		CVE-2022-25845 & 30    & 0     & 2     & 0    & 0      & 2   \\
		CVE-2022-38749 & 21    & 0     & 21    & 1    & 0      & 21  \\
		CVE-2022-38751 & 21    & 0     & 21    & 1    & 0      & 21  \\
		CVE-2022-42889 & 4     & 0     & 4     & 0    & 0      & 4   \\
		CVE-2022-45688 & 419   & 0     & 34    & 1    & 0      & 34  \\ \hline
		all       & 727   & 10    & 137   & 5    & 41     & 149 \\ \hline
	\end{tabular}
	\caption{CVEs detected by various SCA tools in clones found by our approach}
	\label{tab:sca:clones}
\end{table}

\subsection{Scalability}

Experiments were conducted on a server running Linux 6.3.8-arch1-1 with  8 Intel(R) Xeon(R) CPU E5-2637 v4 @ 3.50GHz CPUs, and 64GB RAM. The serial run of the analyses for all 29 CVEs took 12 hours 15 minutes.  We have invested significant effort on caching to improve the scalability of re-running experiments. This includes the caching of both REST query results as well as build results.  This has resulted in re-runs to be faster by order of a magnitude (50 minutes when running the experiments using 4 parallel tasks).

\section{Limitations and Threats to Validity}
\label{sec:threats}

\subsection {Precision}


The analysis presented here  is designed to be precise. This is ensured by making vulnerabilities testable through POVs. However, there is a possibility that those tests do not correctly reflect the vulnerability. Sometimes vulnerabilities are reported in great detail. An example are parser vulnerabilities discovered by fuzzers like \textit{oss-fuzz}~\cite{serebryany2017oss}, which discovers and reports payloads~\footnote{For instance, see \url{https://www.cvedetails.com/cve/CVE-2022-38750/}, \url{https://bitbucket.org/snakeyaml/snakeyaml/issues/526/stackoverflow-oss-fuzz-47027} for a CVE reported by \textit{oss-fuzz}}.  Sometimes, reports are vague (and sometimes this is on purpose as part of the disclosure process), and POVs are constructed from the understanding of an individual programmer of the vulnerability. Sometimes, those tests may miss some additional security measures clones may introduce - for instance, the tests for CVE-2022-42889 in \textit{commons-text:1.9 } check whether interpolator lookup provides entries for the \textit{script}, \textit{dns} and \textit{url} prefixes, and test the execution of an OS command using the script prefix. But the Tests do not check whether actual network lookups happen for those prefixes. This is an engineering compromise -- additional network connectivity makes tests flaky , and slows down the pipeline, and we deem the overall risk that this introduces false positives very low.

\subsection {Soundness}

Our analysis is unsound. As with all program analysis, we have to strike a reasonable balance between precision, scalability and recall, with theoretical and practical limitations implying that a non-trivial analysis that is precise, sound and fast is not possible. Priority was given to precision in line with industry best practices, driven by developer acceptance~\cite{bessey2010few, sadowski2018lessons, distefano2019scaling}.  Scalability considerations had to be taken into account as repositories are very large and evolving, and maintaining a copy is not feasible for economic reasons. Therefore, we have made decisions to limit interactions with  the Maven repositories via the REST API by limiting the number of queries. While some of this can be achieved by engineering (in particular, our tool extensively uses caching, similar to what other Maven clients do), sometimes those restrictions (number of classes used to detect clone candidates, number of results and pages fetched for each query) imply that results are missed. 

Our analysis will also miss clones that are on the subclass level (e.g., single functions), or clones that have custom modifications of source code beyond package renaming and altering or removing comments. Lowering the threshold for clone detection would be interesting, the question being whether this still would lead to the detection of vulnerable artifacts.  This is an area for future research. We expect that  the law of diminishing returns will apply here. 

We think that the proposed simple analysis is still useful as its purpose is not to measure the number of artifacts associated with vulnerabilities, but to demonstrate that this is a significant problem that deserves attention.

Another limitation of our analysis is that it relies on source code. This means that components written in other languages that can be compiled into Java bytecode and deployed in the Maven repository are not covered, and this decreases the detection rate of our tool. The pipeline analysis (Table \ref{tab:pipelinestats}) suggests that for the particular analysis discussed here, this is not a big problem. This problem can be addressed in future work by writing a source-code based clone analysis for alternative languages like Kotlin, or by switching to a bytecode-based method that can abstract from compiler specifics~\cite{dann2019sootdiff} and deal with the effects of non-determinic compilation~\cite{xiong2022towards}. 

\begin{figure}
	\includegraphics[width=6cm]{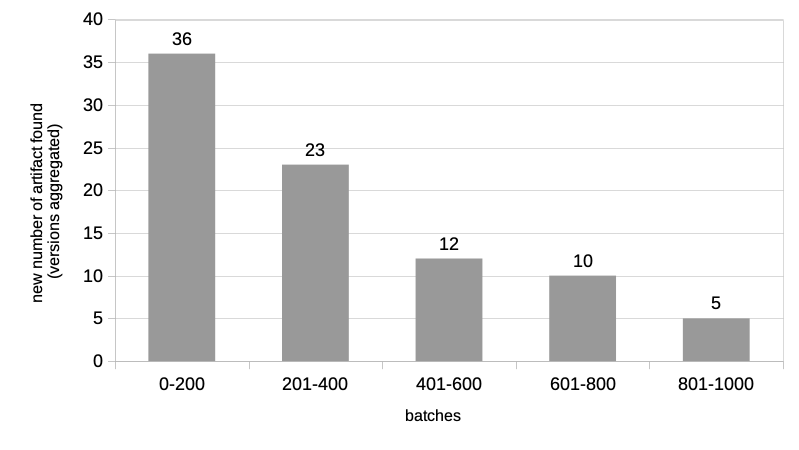}
	\caption{New vulnerable clones retrieved by batch, aggregated (versions ignored)}
	\label{fig:batches}
\end{figure}

We make no claim that the various parameters used in our analysis are optimal. The most obvious way to improve recall is to fetch more data. We had to use paging (page size 200) for queries, and there is some evidence that returns are diminishing  in later batches as shown in Figure~\ref{fig:batches}. The choice of 1000 as initial query size was selected as a good trade-off between performance and recall. 

\subsection {Reproducibility}

By design, there are certain limitations on  reproducibility. Both the repository and the vulnerability database (and therefore the SCA tools) permanently evolve and we expect that many of the vulnerable components we detect will be marked as such eventually as we release results as described in Section~\ref{sec:disclosure}.  

We report the results of running the SCA tools at the time when the experiments were conducted on the artifacts in a release repository (\url{https://github.com/jensdietrich/xshady-release}).  The code containing the actual tool will 
be released following the disclosure to vendor delay.

\section{Disclosure}
\label{sec:disclosure}

\subsection{Disclosure Process}
\label{sec:disclosure:process}

We describe the process we are using to disclose our findings. This is not straight-forward as we are not finding new vulnerabilities, so the standard vulnerability disclosure process does not necessarily apply. Instead, we detect new propagation pathways along which vulnerabilities spread, i.e. hidden dependencies not being detected by existing SCA tools due to their current limitations. 

However, there is a grey zone between cloning or shading a library, and inlining some code that becomes part of a unique new product, with its own unique vulnerabilities. To decide how to disclose the presence of a vulnerability detected, we took the following criteria into account: 

\begin{enumerate}
	\item Whether the project is designed to be a full clone of the original artifact. This is determined by the artifact name being the same or very similar to the name of the original artifact. This can still be the case if the artifact uses shading.
     \item Whether the project is critical.  This is defined by having a low number of contributors to  the associated repository, and no external dependents on Maven central outside the group of the artifact. 
     \item Whether the project has been remediated, interpreted as whether there was a newer version available in the repository at the time of the analysis, and the analysis did not detect the vulnerability in this version.  
\end{enumerate}

Based on this we used a disclosure procedure that has two possible outcomes: database disclosure, or disclosure to vendor. 

For database disclosure, we use a release repository on GitHub~\footnote{The release repository is \url{https://github.com/jensdietrich/xshady-release}, the original POV projects used as input can be found in \url{https://github.com/jensdietrich/xshady}} where we release the instantiated POV projects, and publish results by modifying the entries in the GitHub advisory database via pull requests. 

\begin{figure}
	\includegraphics[width=3cm]{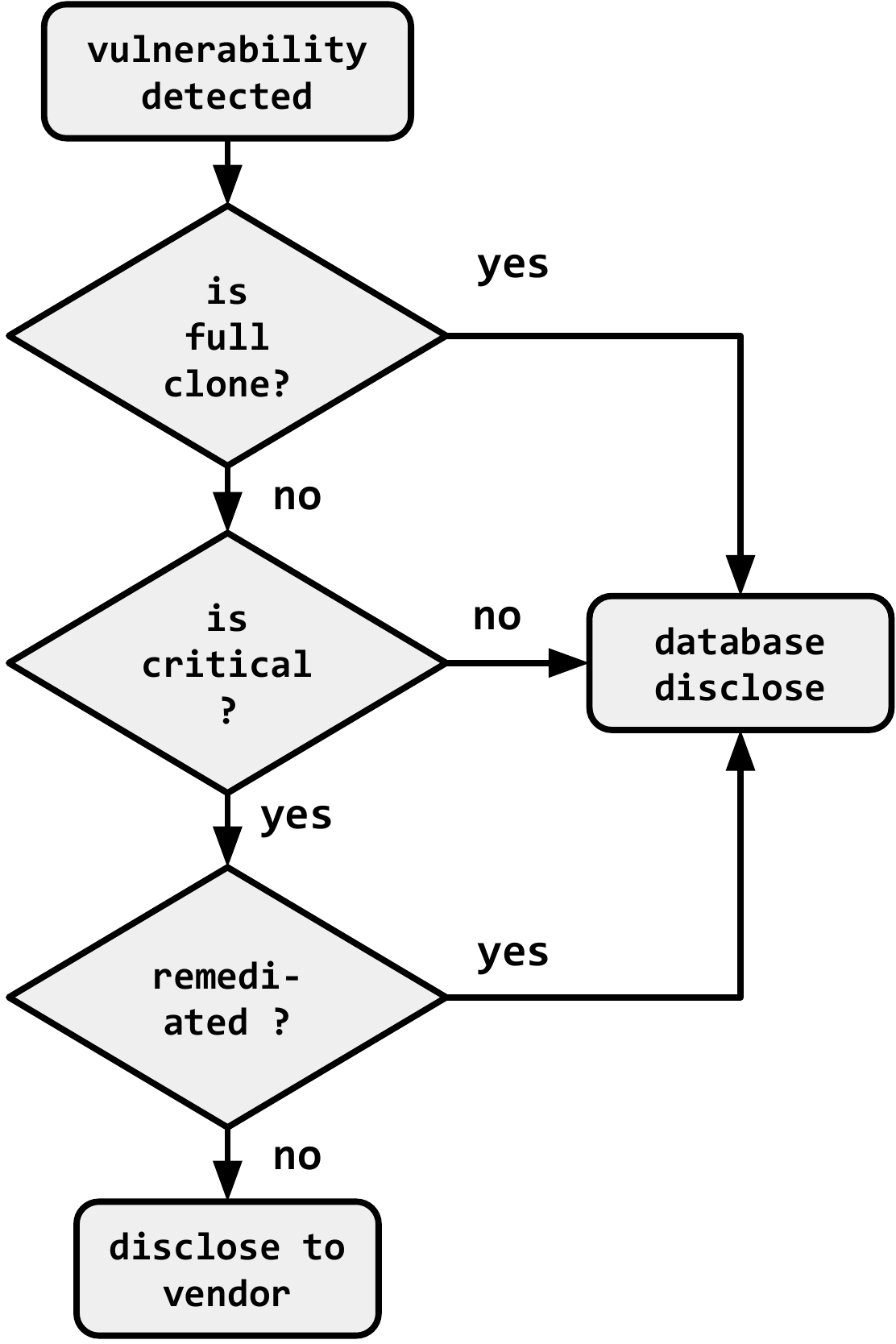}
	\caption{Disclosure procedure}
	\label{fig:disclosure}
\end{figure}

Our disclosure process is depicted in Figure~\ref{fig:disclosure}.

\subsection{Accepted Disclosures}
\label{sec:disclosure:accepted}

At the time of submission, our work had resulted in 6 changes to the GitHub security advisory via accepted pull requests~\footnote{The URL pattern for the respective pull request is \url{https://github.com/github/advisory-database/pull/<id>}}:  CVE-2022-38749 (PR: 2258), CVE-2022-42889 (PR: 2273), CVE-2015-6420 (PR: 2326),  CVE-2016-2510 (PR: 2327), CVE-2018-10237 (PR: 2444),  CVE-2021-44228 (PR: 2445).

We found CVE-2022-45688  in a shaded version of \textit{json.org} in several components in the \textit{org.graalvm.tools} group. Those were disclosed to the vendor, and a patch was announced in the Oracle Critical Patch Update Advisory July 2023~\footnote{\url{https://www.oracle.com/security-alerts/cpujul2023.html}}.  Those vulnerabilities were classified as non-exploitable. 

Future / open GHSA pull requests can be found using the following URL:
\url{https://bit.ly/xshady-ghsa-pr}.

\section{Conclusion}
\label{sec:conclusion}

We have presented a novel lightweight approach to detect the presence of vulnerabilities in components that use cloning and shading. We demonstrated that this reveals blind spots in vulnerability databases and tools relying on those. This is a common problem -- we detected vulnerable clones for more than half of the vulnerabilities studied, including vulnerabilities that are critical, and have been known for years. 

Our results indicate that we need to design software composition analysis tools that perform deeper analyses that do not only rely on project meta-data. Several accepted GHSA pull requests emphasise the practical relevance of our findings.

\section{Acknowledgements}

The authors would like to thank Dhanushka Jayasuriya and Emanuel Evans. The work of the first author was supported by a gift by Oracle Labs Australia, and by the Veracity project funded by the New Zealand National Science Challenge for Technological Innovation (Sfti).


\bibliographystyle{plain}
\bibliography{bibliography}

\begin{thebibliography}{10}

\bibitem{sonartype-plugin}
Apache maven plugin for sonatype oss index.
\newblock \url{https://sonatype.github.io/ossindex-maven/maven-plugin/}.

\bibitem{dependabot}
Dependabot -- automated dependency updates built into github.
\newblock \url{https://github.com/dependabot}.

\bibitem{ghsa}
Github advisory database.
\newblock \url{https://github.com/advisories}.

\bibitem{nvd}
National vulnerability database.
\newblock \url{https://nvd.nist.gov/vuln}.

\bibitem{owasp-dependency-check}
Owasp dependency-check.
\newblock \url{https://owasp.org/www-project-dependency-check/}.

\bibitem{snyk}
snyk.
\newblock \url{https://snyk.io/}.

\bibitem{owaspTop10A06}
A06:2021 – vulnerable and outdated components, 2021.
\newblock
  \url{https://owasp.org/Top10/A06_2021-Vulnerable_and_Outdated_Components/}.

\bibitem{alfadel2023empirical}
Mahmoud Alfadel, Diego~Elias Costa, and Emad Shihab.
\newblock Empirical analysis of security vulnerabilities in python packages.
\newblock {\em Empirical Software Engineering}, 28(3):59, 2023.

\bibitem{baltes2020code}
Sebastian Baltes and Christoph Treude.
\newblock Code duplication on stack overflow.
\newblock In {\em Proceedings of the ACM/IEEE 42nd International Conference on
  Software Engineering: New Ideas and Emerging Results}, pages 13--16, 2020.

\bibitem{bessey2010few}
Al~Bessey, Ken Block, Ben Chelf, Andy Chou, Bryan Fulton, Seth Hallem, Charles
  Henri-Gros, Asya Kamsky, Scott McPeak, and Dawson Engler.
\newblock A few billion lines of code later: using static analysis to find bugs
  in the real world.
\newblock {\em Communications of the ACM}, 53(2):66--75, 2010.

\bibitem{bui2022vul4j}
Quang-Cuong Bui, Riccardo Scandariato, and Nicol{\'a}s E~D{\'\i}az Ferreyra.
\newblock Vul4j: a dataset of reproducible java vulnerabilities geared towards
  the study of program repair techniques.
\newblock In {\em Proceedings of the 19th International Conference on Mining
  Software Repositories}, pages 464--468, 2022.

\bibitem{chowdhury2021untriviality}
Md~Atique~Reza Chowdhury, Rabe Abdalkareem, Emad Shihab, and Bram Adams.
\newblock On the untriviality of trivial packages: An empirical study of npm
  javascript packages.
\newblock {\em IEEE Transactions on Software Engineering}, 48(8):2695--2708,
  2021.

\bibitem{dann2019sootdiff}
Andreas Dann, Ben Hermann, and Eric Bodden.
\newblock Sootdiff: Bytecode comparison across different java compilers.
\newblock In {\em Proceedings of the 8th ACM SIGPLAN International Workshop on
  State of the Art in Program Analysis}, pages 14--19, 2019.

\bibitem{dann2021identifying}
Andreas Dann, Henrik Plate, Ben Hermann, Serena~Elisa Ponta, and Eric Bodden.
\newblock Identifying challenges for oss vulnerability scanners-a study \& test
  suite.
\newblock {\em IEEE Transactions on Software Engineering}, 48(9):3613--3625,
  2021.

\bibitem{decan2018impact}
Alexandre Decan, Tom Mens, and Eleni Constantinou.
\newblock On the impact of security vulnerabilities in the npm package
  dependency network.
\newblock In {\em Proceedings of the 15th international conference on mining
  software repositories}, pages 181--191, 2018.

\bibitem{decan2019empirical}
Alexandre Decan, Tom Mens, and Philippe Grosjean.
\newblock An empirical comparison of dependency network evolution in seven
  software packaging ecosystems.
\newblock {\em Empirical Software Engineering}, 24:381--416, 2019.

\bibitem{dick2018dll}
Stephanie Dick and Daniel Volmar.
\newblock Dll hell: Software dependencies, failure, and the maintenance of
  microsoft windows.
\newblock {\em IEEE Annals of the History of Computing}, 40(4):28--51, 2018.

\bibitem{dietrich2016java}
Jens Dietrich, Kamil Jezek, and Premek Brada.
\newblock What java developers know about compatibility, and why this matters.
\newblock {\em Empirical Software Engineering}, 21:1371--1396, 2016.

\bibitem{distefano2019scaling}
Dino Distefano, Manuel F{\"a}hndrich, Francesco Logozzo, and Peter~W O'Hearn.
\newblock Scaling static analyses at facebook.
\newblock {\em Communications of the ACM}, 62(8):62--70, 2019.

\bibitem{eichberg2014software}
Michael Eichberg and Ben Hermann.
\newblock A software product line for static analyses: the opal framework.
\newblock In {\em Proceedings of the 3rd ACM SIGPLAN International Workshop on
  the State of the Art in Java Program Analysis}, pages 1--6, 2014.

\bibitem{ernst2003static}
Michael~D Ernst.
\newblock Static and dynamic analysis: Synergy and duality.
\newblock In {\em WODA 2003: ICSE Workshop on Dynamic Analysis}, pages 24--27,
  2003.

\bibitem{grove1997call}
David Grove, Greg DeFouw, Jeffrey Dean, and Craig Chambers.
\newblock Call graph construction in object-oriented languages.
\newblock In {\em Proceedings of the 12th ACM SIGPLAN conference on
  Object-oriented programming, systems, languages, and applications}, pages
  108--124, 1997.

\bibitem{Haq2021}
Irfan~Ul Haq and Juan Caballero.
\newblock A survey of binary code similarity.
\newblock {\em ACM Computing Surveys}, 54:1--38, 6 2021.

\bibitem{hiesgen2022race}
Raphael Hiesgen, Marcin Nawrocki, Thomas~C Schmidt, and Matthias W{\"a}hlisch.
\newblock The race to the vulnerable: Measuring the log4j shell incident.
\newblock {\em arXiv preprint arXiv:2205.02544}, 2022.

\bibitem{iannone2021toward}
Emanuele Iannone, Dario Di~Nucci, Antonino Sabetta, and Andrea De~Lucia.
\newblock Toward automated exploit generation for known vulnerabilities in
  open-source libraries.
\newblock In {\em 2021 IEEE/ACM 29th International Conference on Program
  Comprehension (ICPC)}, pages 396--400. IEEE, 2021.

\bibitem{jezek2015java}
Kamil Jezek, Jens Dietrich, and Premek Brada.
\newblock How java apis break--an empirical study.
\newblock {\em Information and Software Technology}, 65:129--146, 2015.

\bibitem{kikas2017structure}
Riivo Kikas, Georgios Gousios, Marlon Dumas, and Dietmar Pfahl.
\newblock Structure and evolution of package dependency networks.
\newblock In {\em 2017 IEEE/ACM 14th International Conference on Mining
  Software Repositories (MSR)}, pages 102--112. IEEE, 2017.

\bibitem{kula2018developers}
Raula~Gaikovina Kula, Daniel~M German, Ali Ouni, Takashi Ishio, and Katsuro
  Inoue.
\newblock Do developers update their library dependencies? an empirical study
  on the impact of security advisories on library migration.
\newblock {\em Empirical Software Engineering}, 23:384--417, 2018.

\bibitem{livshits2015defense}
Benjamin Livshits, Manu Sridharan, Yannis Smaragdakis, Ond{\v{r}}ej Lhot{\'a}k,
  J~Nelson Amaral, Bor-Yuh~Evan Chang, Samuel~Z Guyer, Uday~P Khedker, Anders
  M{\o}ller, and Dimitrios Vardoulakis.
\newblock In defense of soundiness: A manifesto.
\newblock {\em Communications of the ACM}, 58(2):44--46, 2015.

\bibitem{luszcz2018apache}
Jeff Luszcz.
\newblock Apache struts 2: how technical and development gaps caused the
  equifax breach.
\newblock {\em Network Security}, 2018(1):5--8, 2018.

\bibitem{mir2023effect}
Amir~M Mir, Mehdi Keshani, and Sebastian Proksch.
\newblock On the effect of transitivity and granularity on vulnerability
  propagation in the maven ecosystem.
\newblock {\em arXiv preprint arXiv:2301.07972}, 2023.

\bibitem{mirhosseini2017can}
Samim Mirhosseini and Chris Parnin.
\newblock Can automated pull requests encourage software developers to upgrade
  out-of-date dependencies?
\newblock In {\em 2017 32nd IEEE/ACM international conference on automated
  software engineering (ASE)}, pages 84--94. IEEE, 2017.

\bibitem{peng2023impact}
Sida Peng, Eirini Kalliamvakou, Peter Cihon, and Mert Demirer.
\newblock The impact of ai on developer productivity: Evidence from github
  copilot.
\newblock {\em arXiv preprint arXiv:2302.06590}, 2023.

\bibitem{ponta2018beyond}
Serena~Elisa Ponta, Henrik Plate, and Antonino Sabetta.
\newblock Beyond metadata: Code-centric and usage-based analysis of known
  vulnerabilities in open-source software.
\newblock In {\em 2018 IEEE International Conference on Software Maintenance
  and Evolution (ICSME)}, pages 449--460. IEEE, 2018.

\bibitem{raemaekers2014semantic}
Steven Raemaekers, Arie Van~Deursen, and Joost Visser.
\newblock Semantic versioning versus breaking changes: A study of the maven
  repository.
\newblock In {\em 2014 IEEE 14th International Working Conference on Source
  Code Analysis and Manipulation}, pages 215--224. IEEE, 2014.

\bibitem{ragkhitwetsagul2019toxic}
Chaiyong Ragkhitwetsagul, Jens Krinke, Matheus Paixao, Giuseppe Bianco, and
  Rocco Oliveto.
\newblock Toxic code snippets on stack overflow.
\newblock {\em IEEE Transactions on Software Engineering}, 47(3):560--581,
  2019.

\bibitem{rattan2013clones}
Dhavleesh Rattan, Rajesh Bhatia, and Maninder Singh.
\newblock Software clone detection: A systematic review.
\newblock {\em Information and Software Technology}, 55(7):1165--1199, 2013.

\bibitem{rice1953classes}
Henry~Gordon Rice.
\newblock Classes of recursively enumerable sets and their decision problems.
\newblock {\em Transactions of the American Mathematical society},
  74(2):358--366, 1953.

\bibitem{riehle2019open}
Dirk Riehle and Nikolay Harutyunyan.
\newblock Open-source license compliance in software supply chains.
\newblock In {\em Towards Engineering Free/Libre Open Source Software (FLOSS)
  Ecosystems for Impact and Sustainability: Communications of NII Shonan
  Meetings}, pages 83--95. Springer, 2019.

\bibitem{roy2009comparison}
Chanchal~K Roy, James~R Cordy, and Rainer Koschke.
\newblock Comparison and evaluation of code clone detection techniques and
  tools: A qualitative approach.
\newblock {\em Science of computer programming}, 74(7):470--495, 2009.

\bibitem{sadowski2018lessons}
Caitlin Sadowski, Edward Aftandilian, Alex Eagle, Liam Miller-Cushon, and Ciera
  Jaspan.
\newblock Lessons from building static analysis tools at google.
\newblock {\em Communications of the ACM}, 61(4):58--66, 2018.

\bibitem{serebryany2017oss}
Kostya Serebryany.
\newblock Oss-fuzz-google’s continuous fuzzing service for open source
  software.
\newblock In {\em USENIX Security symposium}. USENIX Association, 2017.

\bibitem{sui2020recall}
Li~Sui, Jens Dietrich, Amjed Tahir, and George Fourtounis.
\newblock On the recall of static call graph construction in practice.
\newblock In {\em Proceedings of the ACM/IEEE 42nd International Conference on
  Software Engineering (ICSE'20)}, pages 1049--1060, 2020.

\bibitem{taylor2020defending}
Matthew Taylor, Ruturaj Vaidya, Drew Davidson, Lorenzo De~Carli, and Vaibhav
  Rastogi.
\newblock Defending against package typosquatting.
\newblock In {\em Network and System Security: 14th International Conference,
  NSS 2020, Melbourne, VIC, Australia, November 25--27, 2020, Proceedings 14},
  pages 112--131. Springer, 2020.

\bibitem{CVE-2017-5638}
{The MITRE Corporation}.
\newblock Apache struts 2 vulnerability, 2017.
\newblock \url{https://cve.mitre.org/cgi-bin/cvename.cgi?name=CVE-2017-5638}.

\bibitem{CVE-2021-44228}
{The MITRE Corporation}.
\newblock Apache log4j2 vulnerability, 2021.
\newblock \url{https://cve.mitre.org/cgi-bin/cvename.cgi?name=CVE-2021-44228}.

\bibitem{wang2018dependency}
Ying Wang, Ming Wen, Zhenwei Liu, Rongxin Wu, Rui Wang, Bo~Yang, Hai Yu,
  Zhiliang Zhu, and Shing-Chi Cheung.
\newblock Do the dependency conflicts in my project matter?
\newblock In {\em Proceedings of the 2018 26th ACM joint meeting on european
  software engineering conference and symposium on the foundations of software
  engineering (ESEC/FSE'18)}, pages 319--330, 2018.

\bibitem{williams2014unfortunate}
Jeff Williams and Arshan Dabirsiaghi.
\newblock The unfortunate reality of insecure libraries.
\newblock {\em Asp. Secur. Inc}, 2014.
\newblock
  \url{https://cdn2.hubspot.net/hub/203759/file-1100864196-pdf/docs/Contrast_-_Insecure_Libraries_2014.pdf}.

\bibitem{wittern2016look}
Erik Wittern, Philippe Suter, and Shriram Rajagopalan.
\newblock A look at the dynamics of the javascript package ecosystem.
\newblock In {\em Proceedings of the 13th International Conference on Mining
  Software Repositories}, pages 351--361, 2016.

\bibitem{wuunderstanding}
Yulun Wu, Zeliang Yu, Ming Wen, Qiang Li, Deqing Zou, and Hai Jin.
\newblock Understanding the threats of upstream vulnerabilities to downstream
  projects in the maven ecosystem.
\newblock 2023.

\bibitem{xiong2022towards}
Jiawen Xiong, Yong Shi, Boyuan Chen, Filipe~R Cogo, and Zhen~Ming Jiang.
\newblock Towards build verifiability for java-based systems.
\newblock In {\em Proceedings of the 44th International Conference on Software
  Engineering: Software Engineering in Practice}, pages 297--306, 2022.

\bibitem{Zhan2021}
Xian Zhan, Tianming Liu, Yepang Liu, Yang Liu, Li~Li, Haoyu Wang, and Xiapu
  Luo.
\newblock A systematic assessment on android third-party library detection
  tools.
\newblock {\em IEEE Transactions on Software Engineering}, pages 1--1, 2021.

\end{thebibliography}

\end{document}